\documentclass[nojss,noheadings]{jss}

\usepackage[]{graphicx}\usepackage[]{color}
\makeatletter
\def\maxwidth{   \ifdim\Gin@nat@width>\linewidth
    \linewidth
  \else
    \Gin@nat@width
  \fi
}
\makeatother

\definecolor{fgcolor}{rgb}{0.345, 0.345, 0.345}

\usepackage{framed}


\definecolor{shadecolor}{rgb}{.97, .97, .97}
\definecolor{messagecolor}{rgb}{0, 0, 0}
\definecolor{warningcolor}{rgb}{1, 0, 1}
\definecolor{errorcolor}{rgb}{1, 0, 0}
\newenvironment{knitrout}{}{}

\usepackage{amsmath}
\usepackage{amssymb}

\usepackage{hyperref}
\usepackage{enumerate}

\usepackage{tikz}
\usetikzlibrary{positioning,shapes,arrows,decorations.pathreplacing,calc,automata}

\newcommand{\new}[1]{{
#1}}
\newcommand{\nnew}[1]{{
#1}}
\newcommand{\nnnew}[1]{{
#1}}

\title{Generalised Linear Model Trees with Global Additive Effects}

\author{Heidi ~Seibold\\ University of Zurich\\ LMU Munich \And
	Torsten ~Hothorn\\ University of Zurich \And
        Achim ~Zeileis\\ Universit\"at Innsbruck}
\Plainauthor{Heidi Seibold, Torsten Hothorn, Achim Zeileis}

\Address{
  Heidi Seibold, Torsten Hothorn\\
  Department of Biostatistics\\
  Epidemiology, Biostatistics and Prevention Institute\\
  University of Zurich\\
  Hirschengraben 84\\
  CH-8001 Zurich, Switzerland\\

  Heidi Seibold\\
  Institute for Medical Information Processing, Biometry, and Epidemiology\\
  Ludwig-Maximilans-Universit\"at M\"unchen\\
  Marchioninistr. 15\\
  81377 Munich\\

  Achim Zeileis \\
  Department of Statistics \\
  Faculty of Economics and Statistics \\
  Universit\"at Innsbruck \\
  Universit\"atsstr. 15 \\
  A-6020 Innsbruck, Austria \\
}

\Abstract{
Model-based trees are used to find subgroups in data which differ with respect
to model parameters. In some applications it is natural to keep some parameters
fixed globally for all observations while asking if and how other parameters
vary across subgroups.  Existing implementations of model-based trees can
only deal with the scenario where all parameters depend on the subgroups. We
propose partially additive linear model trees (PALM trees) as an extension of
(generalised) linear model trees (LM and GLM trees, respectively), in which the
model parameters are specified a priori to be estimated either globally from all
observations or locally from the observations within the subgroups determined by
the tree. Simulations show that the method has high power for detecting
subgroups in the presence of global effects and reliably recovers the true
parameters.  Furthermore, treatment-subgroup differences are detected in an
empirical application of the method to data from a mathematics exam: the PALM
tree is able to detect a small subgroup of students that had a disadvantage in
an exam with two versions while adjusting for overall ability effects.
}

\usepackage{fancyhdr}
\fancypagestyle{plain}{
\fancyhf{}
\fancyfoot[LO,LE]{\small This is a preprint of an article accepted for
  publication in \textit{Advances in Data Analysis and Classification},
  \doi{10.1007/s11634-018-0342-1}.\\
  Copyright~{\copyright} 2018 Springer-Verlag\\
}
}

\begin{document}

 \maketitle

\section{Introduction}\label{sec:intro}
Model-based recursive partitioning \citep{zeileis_model-based_2008} is used to
partition data into groups that differ in terms of the parameters in the model.
The method can be applied, for example, to find subgroups in a clinical trial
which differ in terms of treatment effect on a health score
\citep[e.g.][]{seibold_model-based_2015} or areas in a city which differ in
terms of the influence of square metres on the rent price.  Sometimes there are
parameters in the model that one wants to fix for all groups, e.g.\ the effect
of smoking on the health outcome in the clinical trial or the effect of
inflation/deflation on rent prices.  This, however, is not possible in
model-based recursive partitioning as described in
\cite{zeileis_model-based_2008}. Here we propose an algorithm called PALM tree
that is similar to model-based recursive partitioning but allows fixing
parameters over all groups, i.e. only some parameters depend on the tree
structure.

There have been several developments in the past years toward the direction of
combining models and trees, where one part of the model follows a tree
structure and one part does not. The Simultaneous Threshold Interaction
Modeling Algorithm \citep[STIMA,][]{dusseldorp_combining_2010} starts off with
a main effects model and adds interactions based on a tree.
\cite{fokkema_detecting_2015} proposed GLMM tree, a method that is similar to
PALM tree, but is used to fix random effects in a generalised linear
mixed-effects model (GLMM) instead of -- as in PALM tree -- further fixed
effects.  Other approaches going in the direction of GLMM tree are RE-EM tree
\citep{sela_re-em_2012} and MERT \citep{hajjem_mixed_2011}.

In the literature on subgroup analyses for the estimation of treatment effects,
special tree-based procedures have been proposed \citep[see,
e.g.][]{doove_comparison_2014}. These methods are commonly used in the analysis
of clinical trials, but are equally relevant in contexts such as marketing
studies evaluating different marketing strategies or studies on website user
behaviour, where users are randomly served one of two website versions (A/B
testing).  \cite{sies_comparing_2016} review some of the methods in a setting
where there are \nnew{some model covariates with fixed parameters across
all subgroups and varying treamtent effect}.  One promising method in this
review is a method by \cite{zhang_estimating_2012} which estimates rules of
optimal treatment for each patient subgroup (optimal treatment regimes).

\nnnew{
The following sections unfold as follows:
In Section~\ref{sec:methods} we will first describe GLMs and GLM trees as the
basics needed for PALM trees and then go into how PALM trees are computed.
}
Furthermore we will show how model-based trees (LM trees, GLM trees and PALM
trees) can be used for finding subgroups with differential treatment effects.
\nnew{
In Section~\ref{sec:sim} we will show the results of a simulation study in
which we compare LM tree, PALM tree, STIMA and the optimal treatment regime
method by \cite{zhang_estimating_2012}.
}
In Section~\ref{sec:aplic} we will apply the PALM tree to data of a mathematics
exam, where the endpoint is performance in the exam, the ``treatment'' is the
student group (early morning or late group) and the known prognostic factor is
the performance in online tests the students participate in during the
semester. Finally we will discuss strengths and limitations of model-based
trees in general and PALM trees in particular.

\section{Methods}\label{sec:methods}

In this section we first describe the basics needed for PALM trees -- GLMs and
GLM trees -- and then introduce PALM trees and how GLMs and GLM trees are used
in the PALM tree algorithm. We focus on GLMs since LMs are a special case of
GLMs.

\subsection{Basics: GLMs and GLM trees}
\subsubsection{GLMs}
GLMs model the expected response $\mu = \mathbb{E}({y})$
given the covariates $\mathbf{x}$. To fix notation we write the GLM as
$
g({\mu}) = \mathbf{x}^\top \boldsymbol{\beta}
$
where $g$ denotes the link function and $\mathbf{x}^\top \boldsymbol{\beta}$
the linear predictor with coefficient vector $\boldsymbol{\beta}$. The
coefficients are estimated by maximising the log-likelihood. The observation-wise
log-likelihood contributions are denoted by $l((y,\mathbf{x})_i, \boldsymbol{\beta})$
with $i=1, \dots, n$ indexing the $i$-th observation and $l$ is defined depending on
the appropriate exponential family chosen for the GLM (Gaussian, Poisson, etc.).

In the following we will make use of two refinements commonly used in GLMs:
(a) interactions and (b) offsets. Interactions are effects combinding two or
more covariates and can be employed to establish subgroup-specific coefficient
vectors in a single model:
\begin{align}\label{eq:sgglm}
g({\mu}) = \mathbf{\tilde{x}}^\top \boldsymbol{\tilde{\beta}}
         = I(\text{subgroup}_1) \cdot
            \mathbf{x}^\top \boldsymbol{\beta}_1 ~+
         I(\text{subgroup}_2) \cdot
            \mathbf{x}^\top \boldsymbol{\beta}_2 ~+
         \dots
\end{align}
where $I(\text{subgroup}_j)$ equals $1$ for observations in the $j$-th subgroup
and $0$ for others. The combined coefficient vector is simply $\boldsymbol{\tilde{\beta}} =
\boldsymbol{\beta}_1^\top, \boldsymbol{\beta}_2^\top, \dots)^\top$

Offsets in GLMs are useful for incorporating additional terms whose effects are
known or fixed into the linear predictor :
\begin{align}\label{eq:offset}
g({\mu}) = \mathbf{x}^\top \boldsymbol{\beta} + \text{offset}.
\end{align}
Thus, the offset behaves like an additional regressor whose coefficient is not
estimated but fixed, e.g.\ to $1$. A prominent example for offsets in GLMs is
the modeling of rates in Poisson regression, where $\text{offset} = 1 \cdot
\log(\text{exposure})$.

\subsubsection{GLM trees}
Tree algorithms generally split the data recursively into disjoint subgroups
(also called nodes) starting from the so-called root node containing all data
and employing certain split points in the so-called split variables. In case
of GLM trees, the idea is to (1) \emph{estimate} the parameters in a GLM using the current sample (starting
with the full data set), (2) \emph{assess} whether the parameters are stable over
the split variables considered, (3) \emph{split} the sample along the variable
associated with the highest parameter instability, (4) \emph{repeat} the previous
steps recursively until some stopping criterion is met (e.g., with respect to the
size of the sample or the instability of the parameters). Various algorithms
have been suggested that can be employed for such GLM-based recursive partitioning,
including GUIDE \citep{loh_regression_2002}, CTree \citep{hothorn_unbiased_2006},
or MOB \citep{zeileis_model-based_2008} where the latter is used subsequently
and explained in more detail in Section~\ref{sec:alg}.
\begin{knitrout}
\definecolor{shadecolor}{rgb}{0.969, 0.969, 0.969}\color{fgcolor}\begin{figure}

{\centering \includegraphics[width=0.7\textwidth]{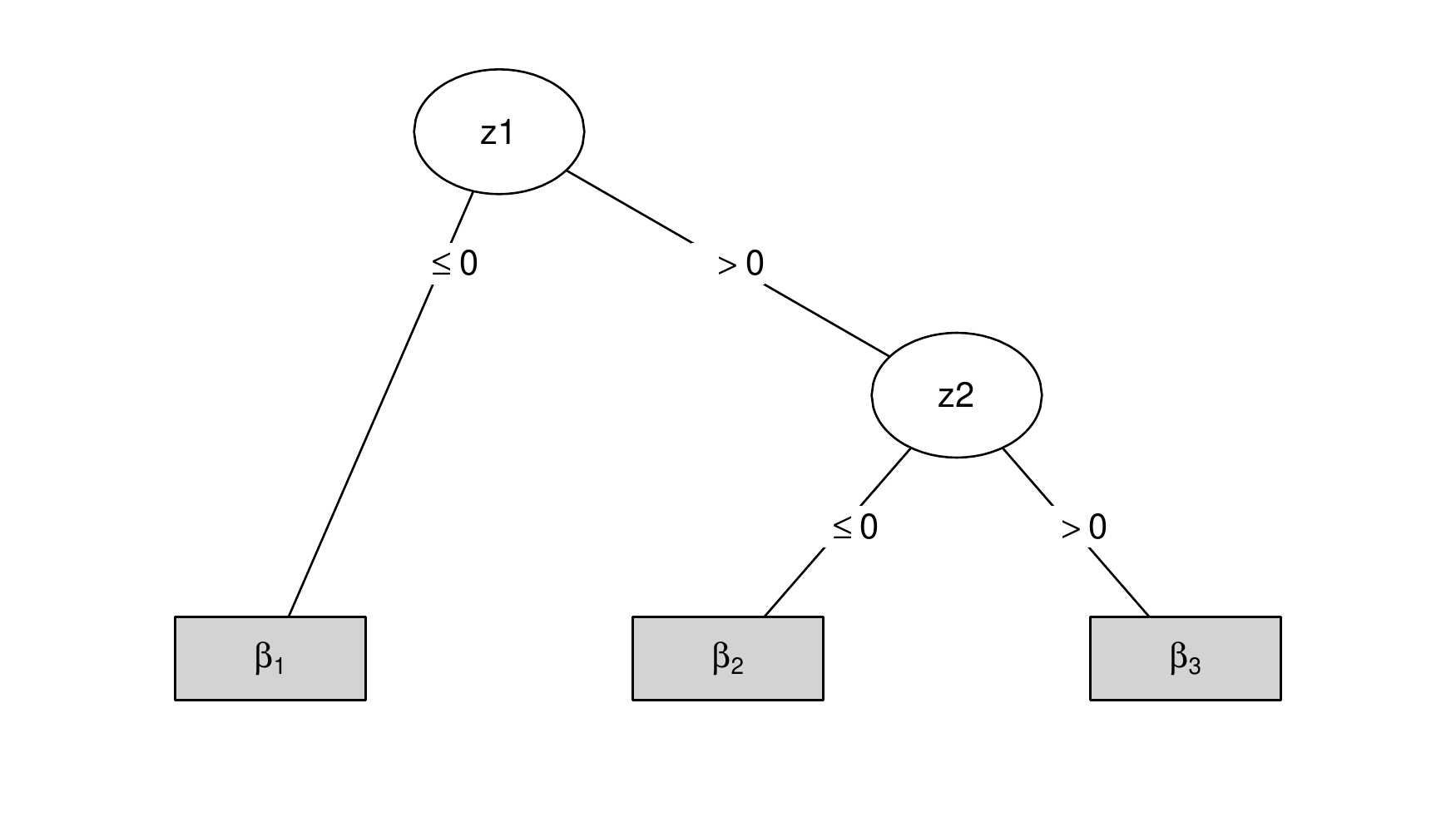}

}

\caption[Example of a model-based tree]{Example of a model-based tree.}\label{fig:sim}
\end{figure}

\end{knitrout}
Figure~\ref{fig:sim} shows an example tree structure that could be found by a
GLM tree with
\begin{align}\label{eq:egbeta}
        \boldsymbol{\beta}(\mathbf{z}) &= \begin{cases}
          \boldsymbol{\beta}_1 & \quad \text{if } ~
                z_1 \leq 0 \\
          \boldsymbol{\beta}_2 & \quad \text{if } ~
                (z_1 > 0) ~\wedge~ (z_2 \leq 0) \\
          \boldsymbol{\beta}_3 & \quad \text{if } ~
                (z_1 > 0) ~\wedge~ (z_2 > 0).
        \end{cases}
\end{align}
The parameters $\boldsymbol{\beta}_1$, $\boldsymbol{\beta}_2$, and $\boldsymbol{\beta}_3$
can be estimated by three separate models for the three subgroups or by using interaction
terms as in Equation~\ref{eq:sgglm} ($I(\text{subgroup}_1) = I(z_1 \leq 0)$
etc.).  To make the role of the split variables more explicit we from now on
write $\mathbf{x}^\top \boldsymbol{\beta}(\mathbf{z})$ instead of
$\mathbf{\tilde{x}}^\top \boldsymbol{\tilde{\beta}}$.
$\boldsymbol{\beta}(\mathbf{z})$ is the interaction effect between covariates
$\mathbf{x}$ and the subgroups defined by the split variables $\mathbf{z}$.

\subsection{Extension: PALM trees}

GLM trees assume that all parameters are subgroup specific. This does not
necessarily have to be the case. PALM trees address this issue and offer a
compromise between GLM trees and GLMs by having one part in which the
parameters depend on subgroups (these are again denoted by
$\boldsymbol{\beta}(\mathbf{z})$) and another part in which the parameters are
the same for all subjects/subgroups (denoted by $\boldsymbol{\gamma}$).

Going from GLMs via GLM trees to PALM trees can be viewed as an evolutionary
process where one method evolves from the other.  The goal of all three is to
appropriately estimate the effect of covariates $\mathbf{x}$ on an outcome
${y}$.  The main difference between the three methods is the structure of the
linear predictor.  While the effects $\boldsymbol{\beta}$ are linear in a GLM,
the effects $\boldsymbol{\beta}(\mathbf{z})$ are linear and constant within
each subgroup but vary between subgroups, i.e. are subgroup-wise linear.
A PALM tree contains globally \textit{fixed} linear effects
$\boldsymbol{\gamma}$ for some covariates $\mathbf{x}_F$ and subgroup-wise
\textit{varying} linear effects $\boldsymbol{\beta}(\mathbf{z})$ for other
covariates $\mathbf{x}_V$.  Mathematically this can be expressed as follows:
\begin{align}
	\text{GLM} \quad & g({\mu}) = \mathbf{x}^\top
		\boldsymbol{\beta}
		\label{mod:basis} \\
	\text{GLM tree} \quad & g({\mu}) = \mathbf{x}^\top
		\boldsymbol{\beta}(\mathbf{z})
		\label{mod:glmtree}\\
	\text{PALM tree} \quad & g({\mu}) = \mathbf{x}_V^\top
		\boldsymbol{\beta}(\mathbf{z}) +
		\mathbf{x}_F^\top \boldsymbol{\gamma}.
		\label{mod:palmtree}
\end{align}
In PALM trees the variables $\mathbf{x}_F$ with a global effect
$\boldsymbol{\gamma}$ have to be defined a priori.  Usually $\mathbf{x}_V$ and
$\mathbf{x}_F$ and $\mathbf{z}$ do not overlap although this is, in principle,
possible.  Note that if the subgroup structure were known,
models~\ref{mod:glmtree} and \ref{mod:palmtree} could both be estimated as GLMs.
Only the fact that it is unknown and has to be detected makes GLM trees and
PALM trees necessary. Also, if the global parameter vector
$\boldsymbol{\gamma}$ were known, model~\ref{mod:palmtree} could be estimated as
GLM tree with $\mathbf{x}_F^\top \boldsymbol{\gamma}$ as offset (as in
equation~\ref{eq:offset}). These connections between the methods are leveraged
in the PALM tree algorithm.

\subsubsection{Algorithm}\label{sec:alg}
We now describe the detailed GLM tree and PALM tree algorithms, starting with
GLM trees as the PALM tree algorithm uses GLM trees in the estimation process.
The GLM tree algorithm is not new and has been explained in depth by
\cite{zeileis_model-based_2008}. The following description of the algorithm
focuses on the parts that are necessary in order to demonstrate the full
concept of the PALM tree algorithm.  Note that to notationally distinguish the
parameters in the subgroups (e.g.\ parameter vector in first subgroup
$\boldsymbol{\beta}_1$) from parameters in the models (e.g.\ first model
parameter $\beta_{(1)}$) we use parentheses.  GLM trees are grown as follows,
starting with the root node containing all observations:
\begin{enumerate}
  \item Compute model~\eqref{mod:basis}, or equivalently
	model~\eqref{mod:glmtree} with a single subgroup
	($\boldsymbol{\beta}(\mathbf{z})= \boldsymbol{\beta}$), in the given node.
  \item Test for instability in the model parameters with respect to each
	of the possible subgroup defining variables
	${Z}_1, \dots, {Z}_J$:
  \begin{itemize}
    \item Compute the score contributions $$s_{(k)}\left((y, \mathbf{x})_i,
    \hat{\boldsymbol{\beta}}\right) = \frac{\partial l\left((y, \mathbf{x})_i,
    \boldsymbol{\beta}\right)}{\partial \beta_{(k)}}
    \bigg\rvert_{\hat{\boldsymbol{\beta}}} $$ as the partial derivatives of the
    log-likelihood contributions of each observation $i$ ($i = 1,\dots,n$) with
    respect to the model parameters $\beta_{(1)}, \dots, \beta_{(K)}$ evaluated
    at the estimated parameters $\hat{\boldsymbol{\beta}} = (\hat{\beta}_{(1)},
    \dots, \hat{\beta}_{(K)})^\top$.
  \item Test if the scores fluctuate randomly around zero for each variable
  $Z_j$ ($j = 1,\dots,J$) , i.e.\ $$H_0^{\beta_{(k)},j}: \quad S_{(k)}\left((Y,
  \mathbf{X}), \hat{\boldsymbol{\beta}}\right) \perp Z_j$$ using M-fluctuation
  tests \citep{zeileis_generalized_2007}.
  \end{itemize}
  \item If the overall test is significant (usually multiplicity adjustment
  using Bonferroni correction is used here), choose variable $Z_j$
  corresponding to the lowest $p$-value as the split variable. In the
  following, we will use $5\%$ as the global significance level.
  \item Choose as split point the point in the split variable which maximizes
  the sum of likelihoods in the emerging subgroups.
  \item Iterate steps 1 to 4 until $H_0^{\beta_{(k)},j} ~\forall k, j$ cannot
  be rejected or some other stop criterion (e.g.\ minimum subgroup size is
  reached) is fulfilled.
\end{enumerate}
The resulting groups differ with respect to at least one of the model
parameters $\boldsymbol{\beta}$. In practice, however, all parameters vary
slightly between subgroups due to the refitting of the model in each node,
i.e.\ for each group of observed subjects. If in reality some covariates
influence the response linearly (for all observations), this leads  to an
overly complex model.  The PALM tree algorithm eliminates this downside by
introducing the possibility to build models where some parameters are kept
stable across subgroups.
This is achieved by starting the estimation of model~(\ref{mod:palmtree}) with
a single subgroup, i.e.\ $\boldsymbol{\beta}(\mathbf{z}) = \boldsymbol{\beta}$,
and then iterating the tree growing process between
\begin{enumerate}[(a)]
 \item estimating $\boldsymbol{\gamma}$ for a given tree structure and
 \item estimating the tree structure for a given $\boldsymbol{\hat{\gamma}}$ (steps 1.-5.).
\end{enumerate}
In (a) we estimate the full model~(\ref{mod:palmtree}) for the known subgroup
$\times$ covariate ($\mathbf{x}_V$) interactions (as in
equation~\ref{eq:sgglm}) and get estimates for $\boldsymbol{\tilde{\beta}}$ and
$\boldsymbol{\gamma}$. In (b) we treat the estimated
$\boldsymbol{\hat{\gamma}}$ as fixed and include $\mathbf{x}_V^\top
\boldsymbol{\hat{\gamma}}$ in the model as an offset.  By preventing
$\boldsymbol{\gamma}$ from being estimated, we exclude it from the score
function and can grow a standard GLM tree (as in steps 1.-5.) for the remaining
parameters.  At the same time we want to account for the effects of
$\mathbf{x}_V$ which is obtained by including $\mathbf{x}_V^\top
\boldsymbol{\hat{\gamma}}$ as offset.  The iterative process stops when no (or
very little) improvement in terms of log-likelihood can be achieved (typically
when the tree structure does not change anymore). Iterating between (a) and (b)
simplifies estimation by only having one unknown: either $\boldsymbol{\gamma}$
or the tree structure.  $\boldsymbol{\beta}(\mathbf{z})$ is estimated in both
steps: In (a) by estimating the model with the known subgroup $\times$
covariate interactions, and in (b) by estimating a separate model for each
subgroup.

PALM trees inherit many of their theoretical properties from the methods
used as building blocks (model-based trees and parametric models), provided
that the model is well specified: Given that the group structure is correctly
detected by the tree, the (G)LM can consistently estimate all coefficients
(grouped and global). Conversely, given that the global coefficients are
estimated consistently, the (G)LM tree uses a group detection based on locally
consistent tests \citep{zeileis_generalized_2007} and the usual locally optimal
greedy forward selection in recursive partitioning \citep[see
e.g.][]{breiman_classification_1984}. To the best of our knowledge, there is no
formal proof that alternating between (a) and (b) will converge to an
``optimal'' solution so that the strengths of both components are guaranteed
to be effective. However, our simulation results (see Section~\ref{sec:sim} and
Appendix~\ref{sec:fullf}) show that PALM trees typically converge quickly and
reliably. This was also found for RE-EM trees \citep{sela_re-em_2012}. While
there is no guarantee that this is always the case, we have not experienced any
convergence issues thus far.

\subsection{Special application: Treatment effects}
One common application of model-based trees is for subgroup analyses in
clinical trials \citep{lipkovich_tutorial_2016, seibold_model-based_2015,
doove_comparison_2014}. In the simplest case one is interested in a treatment
effect of a new treatment versus standard of care or no treatment, i.e.\
$\mathbf{x} \text{ or } \mathbf{x}_V = (1, {x}_A)$ with $x_{Ai} =
I(\text{patient $i$ received new treatment})$. In this setting one
differentiates between prognostic and predictive factors
\citep{italiano_prognostic_2011}. Prognostic factors are patient
characteristics (measured before treatment start) which directly impact the
response, e.g. a health score.  Predictive factors are patient characteristics
which impact the efficacy of the treatment. In the PALM tree framework,
predictive factors should be included in the split variables $\mathbf{z}$ and
prognostic factors, if known in advance, can be included in $\mathbf{x}_F$.
\nnew{
In fact, prognostic factors are often known in advance based on
previous research about the disease.
}

\nnew{
In subgroup analyses for treatment effects the term optimal treatment regime is
commonly mentioned. An optimal treatment regime is a rule which indicates which
treatment is better in which subgroup.
}
Treatment regimes only check the sign of the treatment effect in each subgroup.
If they differ between subgroups, the treatment effects are called qualitative;
if one treatment is better than the other in all subgroups, they are called
quantitative.
\new{
As this application is very common, the remainder of this manuscript will deal
with scenarios where the partitionable parameters are the intercept and the
effect of a binary covariate.
}

\subsection{Comparison to other approaches}\label{sec:other}
\new{
GLMM trees \citep{fokkema_detecting_2015} are closely related to PALM trees, as
the algorithm also builds on the GLM tree algorithm and like PALM tree keeps
parts of the model stable.
}
The major difference is the fact that GLMM trees focus, as the name says, on
generalised mixed effects models and the part that is being kept stable across
subgroups are the random effects.

\new{
STIMA \citep{dusseldorp_combining_2010} is a tree algorithm where the first
split is made in an a priori specified variable, which in the treatment case is
the treatment indicator. All further splits are found by an exhaustive search
and finally a cross-validation based pruning procedure is run to find the
optimal tree. STIMA is similar to PALM tree in the sense that it starts off
with a main effects model and new splits are selected based on a measure of
variance-accounted-for. The main effects of the model are kept stable across
groups and additional effects are added to the model based on the tree
structure. A very similar approach is called partially linear tree-based
regression model \citep[PLTR,][]{chen_partially_2007, cyprien_gpltr_2015}, which
was initially invented to analyse gene-gene and gene-environment effects.
}

The approach by \cite{zhang_estimating_2012} aims to estimate optimal
treatment regimes and is only used in the treatment effect application. In the
following we will use the term OTR (optimal treatment regimes) for this method.
OTR is not as closely related to PALM tree as the previously mentioned methods,
but has shown good performance in settings in which PALM trees are appropriate
\citep{sies_comparing_2016}. OTR does not target estimating the treatment
effect itself but targets learning which treatment is superior for certain
groups of patients.
\new{
OTR starts off with the so-called outcome model, which includes main effects
and treatment $\times$ patient characteristics interactions.
}
After estimating the model the algorithm proceeds as follows:
\new{
\begin{enumerate}
	\item For all patients in the training data predict the response under
	 treatment $\hat{{\mu}}_1$ and under control
	 $\hat{{\mu}}_0$ from the outcome model. Determine the
	 difference $\hat{{\mu}}_1 - \hat{{\mu}}_0$
	 between the two.
	\item Compute a classification algorithm using
	 $I(\hat{{\mu}}_1 - \hat{{\mu}}_0 > 0)$ as
	 response and $|\hat{{\mu}}_1 - \hat{{\mu}}_0|$
	 as weights.
\end{enumerate}
}
Any classification method that can deal with (non-integer) weights could be
used in step 2.

For further tree-based approaches that allow doing analyses similar to
model-based trees see \cite{doove_comparison_2014}.

\section{Simulation study}\label{sec:sim}

\begin{table}
\begin{center}
 \begin{tabular}{rccc}
  \hline
  Simulation variable & Default & Variation & \# Values \\
  \hline
  Difference in treatment effects $\Delta_\beta$       & 0.5 & 0.1--1.5 & 8 \\
  Number of observations $n$                           & 300 & 100--900 & 5 \\
  Qualitative treatment $\times$ subgroup interaction  & Yes & Yes/No   & 2 \\
  Number of patient characteristics $m$                &  30 & 10--70   & 4 \\
  Number of predictive factors $p$                     &   2 & 1--4, 0  & 4, 1 \\
  Number of prognostic factors $q$                     &   2 & 1--4     & 4 \\
  \hline
 \end{tabular}
\end{center}
\caption{Simulation settings. For each scenario one simulation variable is
varied and the rest are kept to the standard value. The value $p=0$ is only
used for the assessment of the type 1 error rate (Section~\ref{sec:simerr1}).}
\label{tab:sim}
\end{table}

\new{
We compare the performance of PALM trees, LM trees, the trees grown based on
the algorithm proposed by \cite{zhang_estimating_2012} (OTR) and STIMA in the
treatment effect setting. We chose OTR as competitor because it showed good
perfomance in scenarios where PALM trees should perform well
\citep{sies_comparing_2016} and we chose STIMA because it is a natural
competitor due to the similarity of the resulting model.  Note that while the
setup of the simulation study is motivated by treatment effect studies, the
insights are of broader interest due to its general structure.
}
The aim is to
evaluate the methods with respect to (1) finding the correct subgroups
(Section~\ref{sec:simsg}), (2) not splitting when there are no subgroups
(Section~\ref{sec:simerr1}), (3) finding the optimal treatment regime
(Section~\ref{sec:simotr}), and (4) correctly estimating the treatment effect
(Section~\ref{sec:simest}). Note that evaluations (1) and (2) are connected
in the sense that they both evaluate the ability to find the correct subgroups.
Furthermore, (3) and (4) are connected in the sense that both evaluate the
ability to give good treatment recommendations.

\new{
We simulate a binary variable (treatment indicator) $X_A$ which is either $1$
or $0$, each with probability $0.5$, and $m$ correlated variables (patient
characteristics)
\begin{align}
	\mathbf{Z} \sim \mathcal{N}_m(\mathbf{0}, \boldsymbol{\Sigma})
\end{align}
with
\begin{align}
        \boldsymbol{\Sigma} =
 \begin{pmatrix}
  1 & 0.2 & \cdots & 0.2 \\
  0.2 & 1 & \cdots & 0.2 \\
  \vdots  & \vdots  & \ddots & \vdots  \\
  0.2 & 0.2 & \cdots & 1
 \end{pmatrix}.
\end{align}
We define the first $p$ variables $Z_1, \dots, Z_p$ to be the true predictive
factors, i.e.\ the patient characteristics that actually interact with the
treatment and thus pose relevant split variables.  The cutpoint is always at
$Z_j = 0$ and the subsequent split is always in the subgroup with $Z_j > 0$,
i.e. on the right side of the tree when visualised as in Figure~\ref{fig:sim}.
We define the consecutive $q$ variables $X_F = (Z_{p+1}, \dots, Z_{p+q})$ to be
the true and known prognostic factors. All further patient characteristics
$Z_{p+q+1}, \dots, Z_m$  are noise variables. We simulate the outcome variable
$Y$ with
\begin{align}
   Y = X_A \beta(\mathbf{Z}) +
      X_F \boldsymbol{\gamma} +
      U
\end{align}
where $U \sim \mathcal{N}(0, 1.5)$ is the error term.

 The effect of the prognostic factors is set to $\boldsymbol{\gamma} =
\mathbf{1}$. The treatment effect $\beta(\mathbf{Z})$ follows a tree structure,
which is visualised in Figure~\ref{fig:sim} for the scenarios with $p=2$.
}
The mathematical representation is as in Equation~\eqref{eq:egbeta} with a
fixed difference between the effects in the subgroups $\Delta_\beta$. We define
a default simulation scenario, which is shown in the second column of
Table~\ref{tab:sim}.  In this default scenario $\Delta_\beta = 0.5$ and
\new{
\begin{align}\label{eq:simbeta}
	\beta(\mathbf{Z}) &= \left\{ \begin{array}{rll}
	  -0.375 &= \beta_1 & \quad \text{if } Z_1 \leq 0 \\
	  ~0.125 &= \beta_2 = \beta_1 + \Delta_\beta &
		\quad \text{if } Z_1 > 0 ~\wedge~ Z_2 \leq 0 \\
	  ~0.625 &= \beta_3 = \beta_2 + \Delta_\beta &
		\quad \text{if } Z_1 > 0 ~\wedge~ Z_2 > 0.
	\end{array} \right.
\end{align}
}
To obtain a diverse set of simulation scenarios which are comparable, we fix
all but one of the simulation variables to the default.  The range of
variation of each simulation variable is given in the third column of
Table~\ref{tab:sim} alongside the number of equidistant values considered (\#
Values). From this we get all necessary information about the simulation, e.g.
$q$ takes 4 different values $1,2,3,4$.  For each distinct simulation setting
we simulate 150
data sets. Note that just for the assessment
of the type 1 error rate (Section~\ref{sec:simerr1}) the number of predictive
factors is set to zero. For the simulation scenarios where $p \neq 2$ and thus
less/more than three true subgroups exist, $\beta(\mathbf{Z})$ follows the same
logic as in Equation~\eqref{eq:simbeta}, i.e. $\beta_b = \beta_{b-1} +
\Delta_\beta$ for $b = 2, \dots, (p+1)$. The value of $\beta_1$ depends on
whether the first split is qualitative or not and on $\Delta_\beta$. If the
first split is not qualitative then $\beta(1) = 0.5$. If the first split is
qualitative $\beta(1) = -3/4 \cdot \Delta_\beta$.  This also means that any
consecutive splits after the first are quantitative.
\new{
This simulation study is limited due to the fact that we only change one
simulation variable at a time.  Section~\ref{sec:fullf} in the Appendix shows
selected results from a full factorial simulation study.
}
Using the simulated data we compare the following methods:
\begin{description}
	\item[PALM tree] with $\mathbf{x}_V = (\mathbf{1}, \mathbf{x}_A)$ and
$\mathbf{x}_F = (z_{p+1}, \dots, z_{p+q})$. The only way we could have
specified this algorithm better for the given data generating process would
have been to add the intercept to $\mathbf{x}_F$, but in real application one
would usually allow the intercept to vary to account for unknown prognostic
factors contained in $\mathbf{z}$.
	\item[LM tree 1] with $\mathbf{x} = (\mathbf{1}, \mathbf{x}_A)$. This
algorithm is of interest to see how well a misspecified model-based tree
behaves. LM tree 1 has to approximate $\mathbf{x}_F^\top \boldsymbol{\gamma}$
using step functions and thus cannot give good results in terms of most
measures used below. However, we are interested in how well it can do in terms
of estimating the correct treatment regime.
	\item[LM tree 2] with $\mathbf{x} = (\mathbf{1}, \mathbf{x}_A,
\mathbf{x}_F)$. This tree is expected to behave better than LM tree 1, since it
contains the correct covariates in the model, but worse than PALM tree since
it may split with respect to instabilities in the parameters for $\mathbf{x}_F$
plus it is overly complex due to the fitting of separate
$\mathbf{x}_F$-parameters in each subgroup.
	\item[OTR] with outcome model $ g(\boldsymbol{\mu}) = (\mathbf{1},
\mathbf{x}_A, \mathbf{x}_F)^\top \boldsymbol{\gamma} + (\mathbf{x}_A :
\mathbf{z})^\top \boldsymbol{\beta}$ (with $\mathbf{x}_A : \mathbf{z}$
interaction between $\mathbf{x}_A$ and $\mathbf{z}$) and pruned CARTs
\citep[Classification and Regression Trees,][]{breiman_classification_1984} as
classification method. OTR was invented to find optimal treatment regimes and
thus is expected to be good at finding the right treatment. OTR is not intended
to find quantitative interactions and thus can not be good at this.
\new{
	\item[STIMA] with a forced first split in the treatment and the maximum
number of splits fixed to six.
}
\end{description}

\subsection{Are the correct subgroups found?}\label{sec:simsg}

\begin{figure}
\includegraphics[width=\maxwidth]{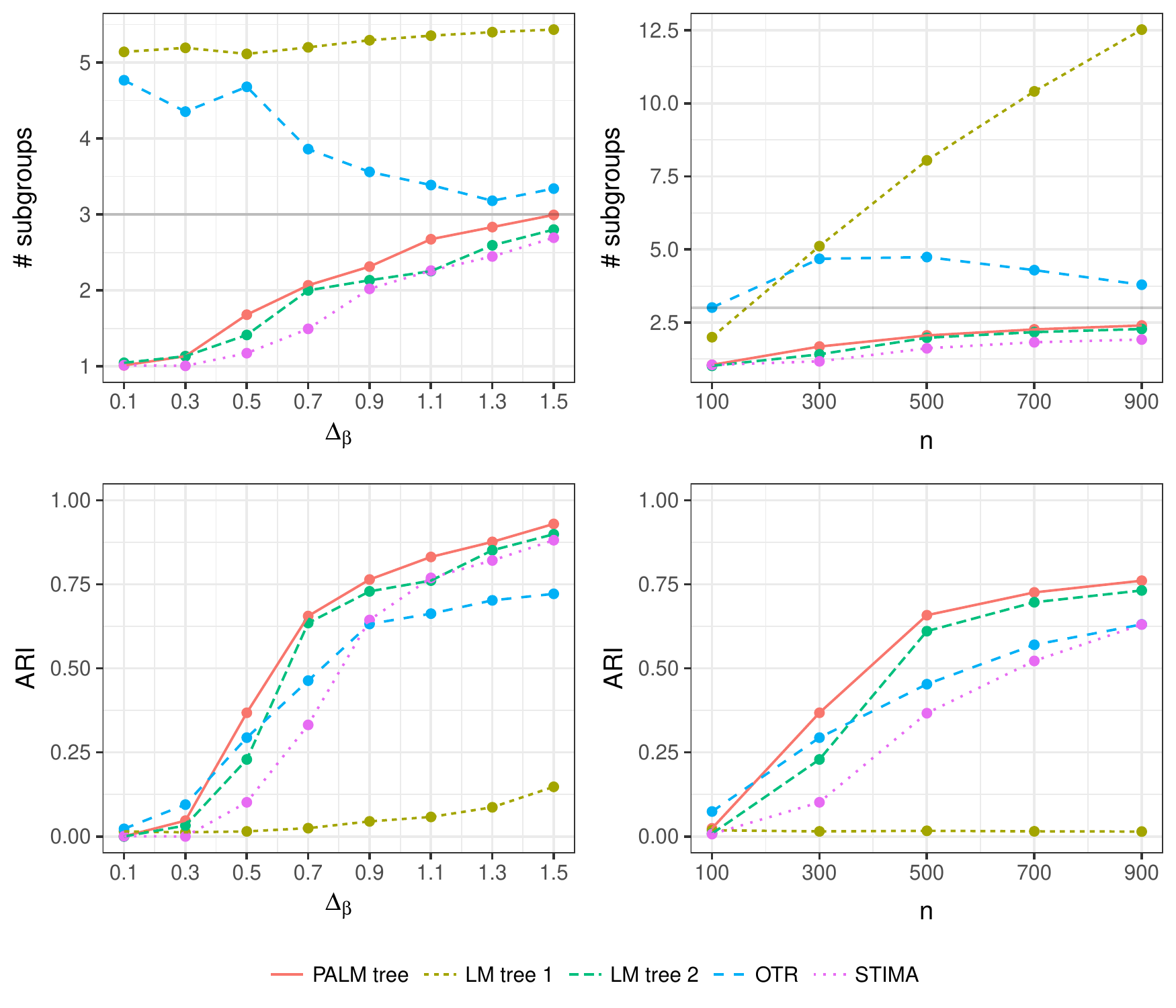}
\caption{Mean number of subgroups and mean ARI for varying $\Delta_\beta$ and number of observations (Question~\ref{sec:simsg}).}\label{fig:fig_nsg1}
\end{figure}


\begin{knitrout}
\definecolor{shadecolor}{rgb}{0.969, 0.969, 0.969}\color{fgcolor}\begin{figure}
\includegraphics[width=\maxwidth]{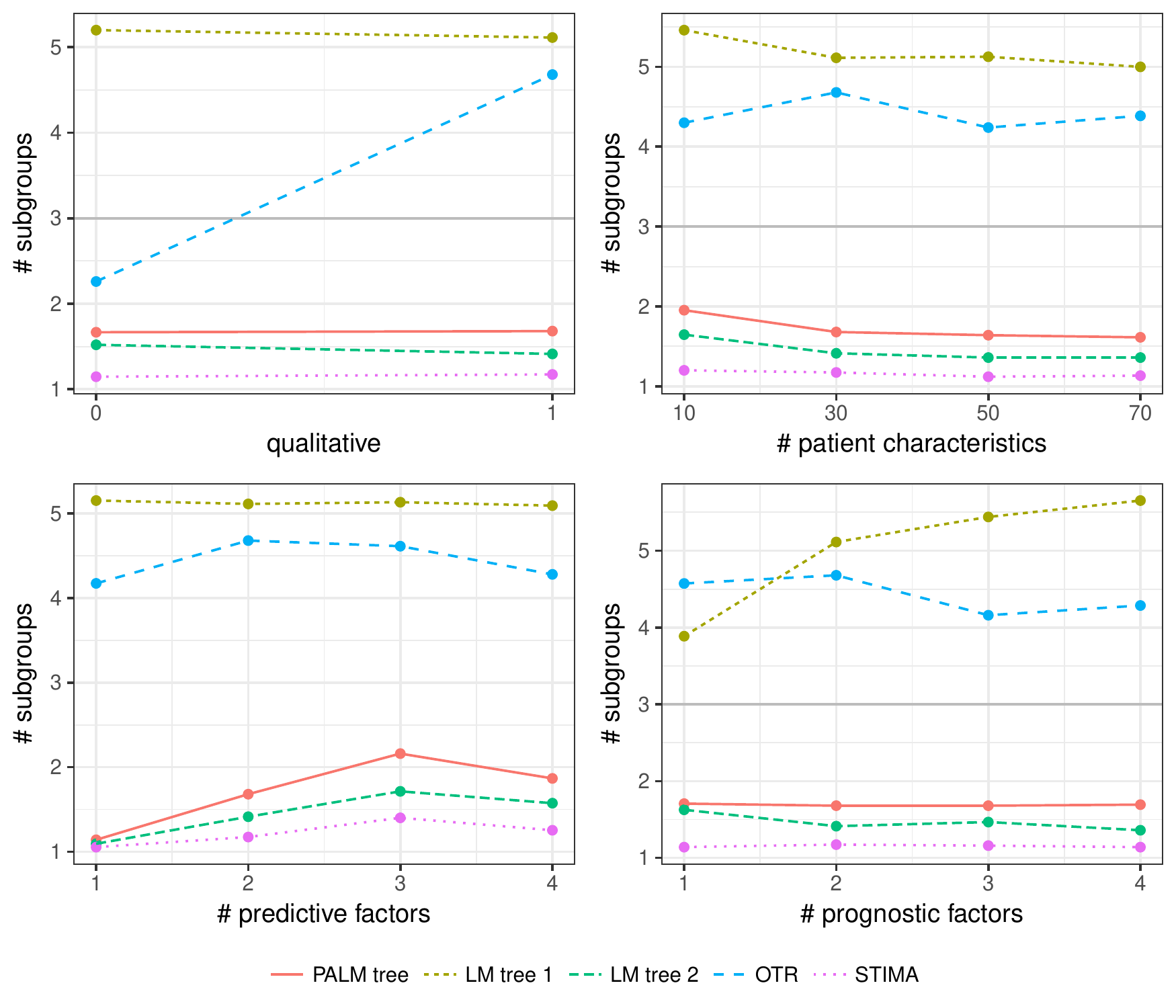}
\caption{Mean number of subgroups for varying types of subgroups (quantitative/qualitative), number of patient characteristics, predictive factors and prognostic factors (Question~\ref{sec:simsg}).}\label{fig:fig_nsg2}
\end{figure}

\end{knitrout}

To investigate whether the correct subgroups are captured by the different
methods, we looked at the number of subgroups found as well as the adjusted Rand
index \citep[ARI,][]{hubert_comparing_1985, milligan_study_1986}. The ARI
measures how well the retrieved subgroups fit with the true underlying
subgroups. If the subgroups found are similar to the true subgroups the ARI
will have a value up to 1. If the subgroups are only as good as a random group
assignment the ARI is 0. If there is systematic missclassification, the ARI can
also be negative.

\new{
The first row of Figure~\ref{fig:fig_nsg1} shows the mean number of selected
subgroups over the 150 simulated data sets and their
corresponding trees for differing \emph{distances between treatment effects
$\Delta_\beta$} and differing \emph{numbers of observations $n$}. This means we
are looking at the case where all variables are kept at the standard value
except $\Delta_\beta$ or $n$ respectively. The second row shows the
corresponding ARI.  The similarity between the PALM tree and LM tree 2
algorithms is obvious. For both the number of subgroups and the ARI the results
are very similar, although PALM tree is slightly better. Both algorithms get
steadily closer to the optimal solution with increasing $\Delta_\beta$ as well
as with increasing number of observations.  LM tree 1 performs badly because it
approximates the linear relation between the prognostic factors and the
response with splits in the data. This is also the reason why with increasing
$n$ the number of subgroups increases. This effect muffles the grouping with
respect to the treatment effect, even if it gets less with increasing
$\Delta_\beta$. The number of subgroups found for OTR is on average greater
than the actual number of subgroups (3 for the given scenarios in
Figure~\ref{fig:fig_nsg1}). The variability of the number of subgroups for OTR
is very high (with a maximum of 20
subgroups).  The true subgroups are not captured as well as with PALM tree and
LM tree 2.  The ARI for OTR is lower than the ARI of PALM tree and LM tree 2
except for very low values of $\Delta_\beta$ and $n$, which can be explained by
the fact that the model-based trees use statistical significance tests and CART
does not.  Even though the pattern of STIMA in terms of the average number of
subgroups appears similar to PALM tree and LM tree 2, on average the ARI is
considerably lower, except for very large differences in treatment effects
($\Delta_\beta$).

Figure~\ref{fig:fig_nsg2} shows the mean number of subgroups for the remaining
simulation scenarios. The model-based trees and STIMA are not affected by the
\emph{type of subgroup}. OTR, however, is designed to find only qualitative
subgroups and thus on average finds fewer groups when there are only
quantitatively differing subgroups. For increasing \emph{number of patient
characteristics}, the model-based trees become more conservative
and find slightly less subgroups, which is due to the correction for multiple
testing (Bonferroni correction). OTR and STIMA do not change much in terms of
average number of subgroups when the number of patient characteristics
increases.  With increasing \emph{number of predictive factors} the number of
subgroups should increase. The true number of subgroups is always the number of
predictive factors $+~1$.  The lower left panel of Figure~\ref{fig:fig_nsg2}
shows that this is not the case for any of the algorithms. The reason for this
is the way of how we simulated the data.  With an increasing number of
predictive factors the subgroups get smaller and thus there is less power to
find splits. The only algorithm that is strongly affected by the \emph{number
of prognostic factors} is LM tree 1, which corresponds to the fact that there
are more linear terms to approximate through the tree structure.
}

\subsection{How often are subgroups found even though there are none?}\label{sec:simerr1}
\begin{knitrout}
\definecolor{shadecolor}{rgb}{0.969, 0.969, 0.969}\color{fgcolor}\begin{figure}

{\centering \includegraphics[width=0.65\textwidth]{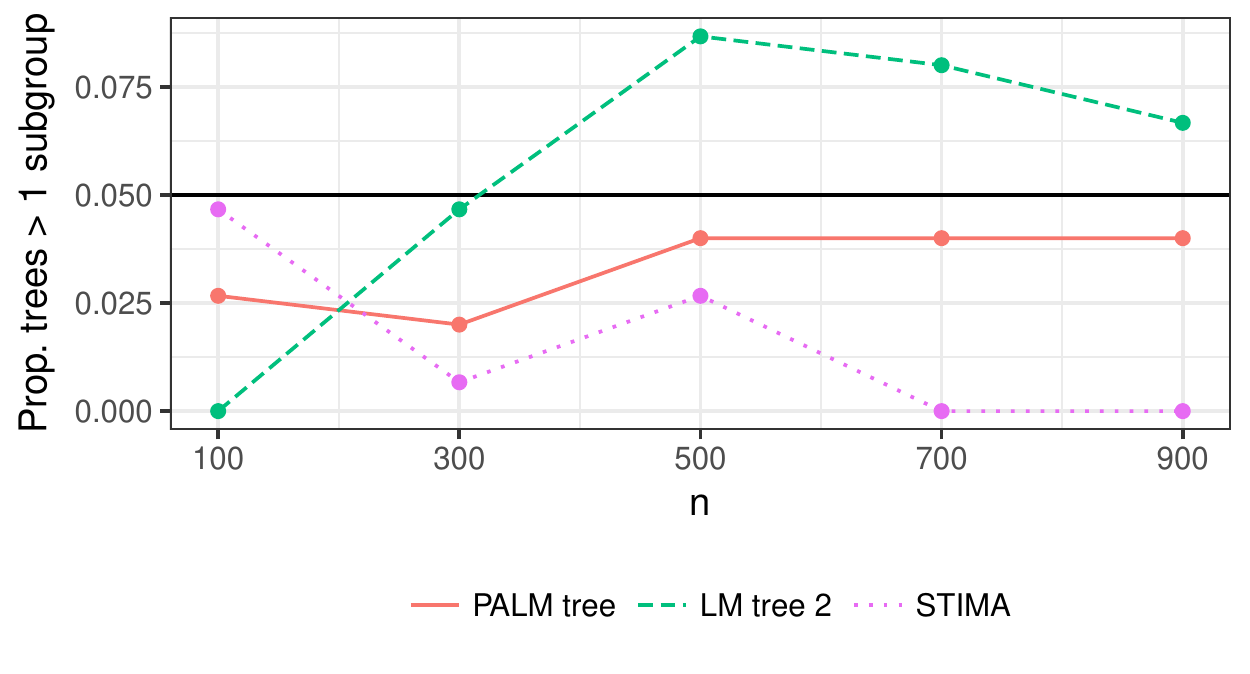}}

\caption{Proportion of trees with more than one subgroup for varying number of observations (Question~\ref{sec:simerr1}). Black line at 0.05.}\label{fig:fig_size}
\end{figure}

\end{knitrout}

To investigate the type 1 error rate, i.e.\ the probability that subgroups are
found even though there are none, we simulated data as above, but with no
predictive factors. This means the treatment effect is the same for all
patients. Figure~\ref{fig:fig_size} shows the behaviour of the methods with
changing \emph{number of observations}.
\new{
LM tree 1 and OTR have a constant value of 1 here and are not visualised. Since
LM tree 1 finds subgroups that have to do with the prognostic factors the
``bad'' performace exists by design. PALM tree is close to the expected $5\%$
significance level, as is LM tree 2. STIMA goes down to $0\%$
for 700 and 900 observations.
}

\subsection{Is the correct treatment predicted to be better?}\label{sec:simotr}
\begin{knitrout}
\definecolor{shadecolor}{rgb}{0.969, 0.969, 0.969}\color{fgcolor}\begin{figure}
\includegraphics[width=\maxwidth]{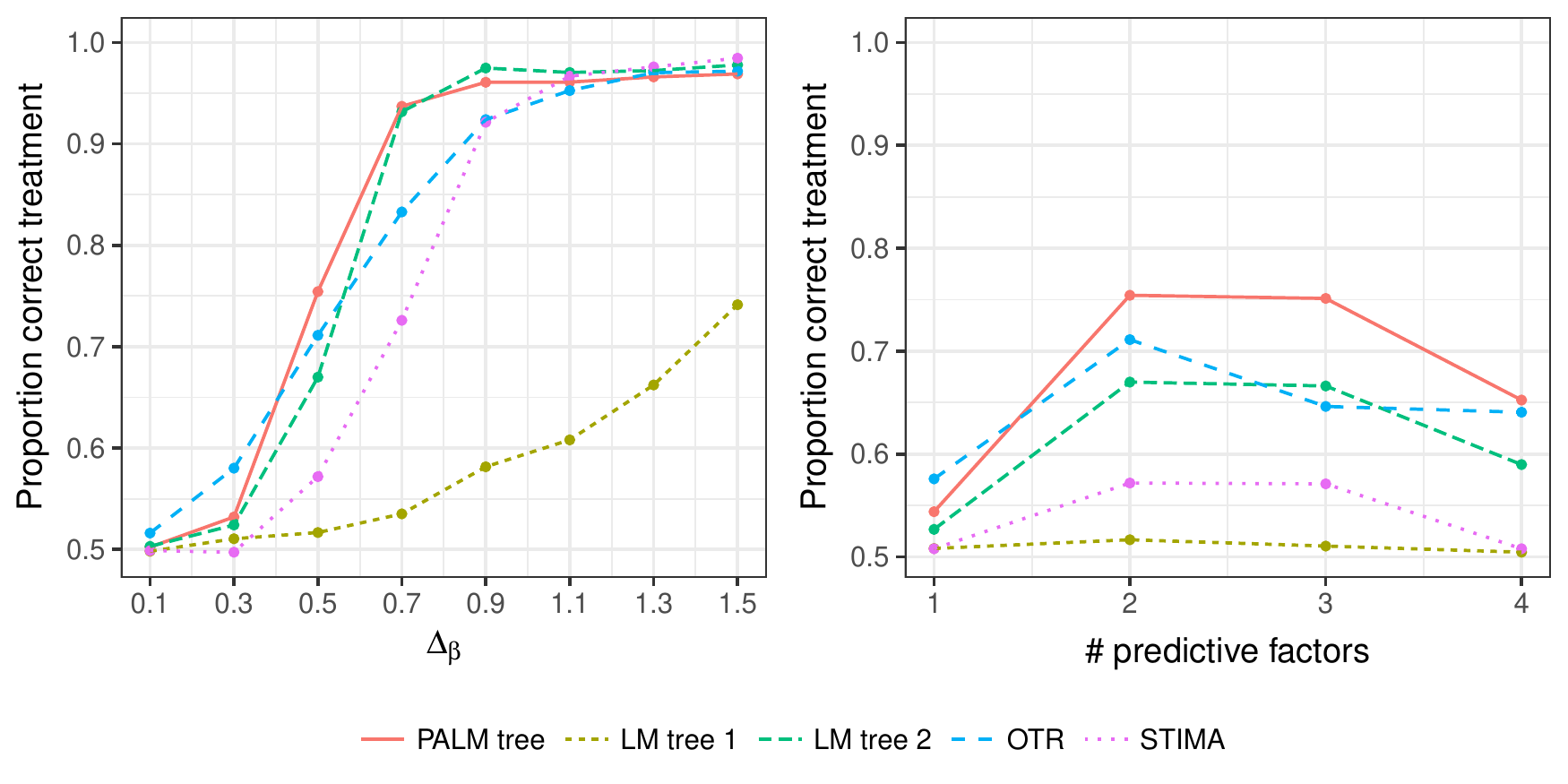}
\caption{Proportion of observations in all trees where better treatment is correctly identified (Question~\ref{sec:simotr}).}\label{fig:fig_cortrt}
\end{figure}

\end{knitrout}

The next measure we wanted to look at is the proportion of patients for which
the better treatment is correctly identified. This is what OTR was designed to
be good at and especially due to the way we simulated data (with a simple
interaction) OTR can be expected to perform well.  Figure~\ref{fig:fig_cortrt}
shows the proportion of patients for which the better treatment is correctly
identified for the scenarios with varying  difference between treatment effects
$\Delta_\beta$ and varying number of predictive factors.  When the
\emph{difference between treatment effects $\Delta_\beta$} is small it is
difficult for all methods to predict the correct treatment regime.  For
$\Delta_\beta = 0.1$ it is close to random guessing. With increasing
$\Delta_\beta$ all methods get better.
\new{
The performance of PALM tree, LM tree 2, OTR and STIMA is similar. The four
methods also behave similarly with a changing \emph{number of predictive
factors}. The treatment regime prediction is globally worst on average when
there is one predictive factor.  This results from the fact that often no split
is found in this case (see Figure~\ref{fig:fig_nsg2}). In cases where the
methods decide not to split at all, this leads by simulation design to a
proportion of $50\%$ correctly-defined treatment regimes.  The proportion of
patients for which the correct treatment is predicted to be the better
treatment improves in cases of two or three predictive factors and gets worse
with four predictive factors.
\nnew{
With more complex and smaller subgroups it becomes more difficult for the
algorithms to retrieve the correct subgroup structure and to estimate the
treatment effect.  Note, however, that shape of the shape of the curves in the
right panel of Figure~\ref{fig:fig_cortrt} is very specific for the simulation
settings here.
}
Figure~\ref{fig:fullf2} shows the results for other scenarios. For example,
for $\Delta_\beta = 1.5$ and 300 observations in a setting with qualitative
treatment differences, the best performace of PALM tree is with only one
predictive factor and decreases from there. The performance of all algorithms
is well in quantitative settings. OTR is the only algorithm that goes down to
only 80\% correctly defined treatment regimes in settings with 100
observations.
}

\subsection{How good is the treatment effect estimate?}\label{sec:simest}

\begin{knitrout}
\definecolor{shadecolor}{rgb}{0.969, 0.969, 0.969}\color{fgcolor}\begin{figure}
\includegraphics[width=\maxwidth]{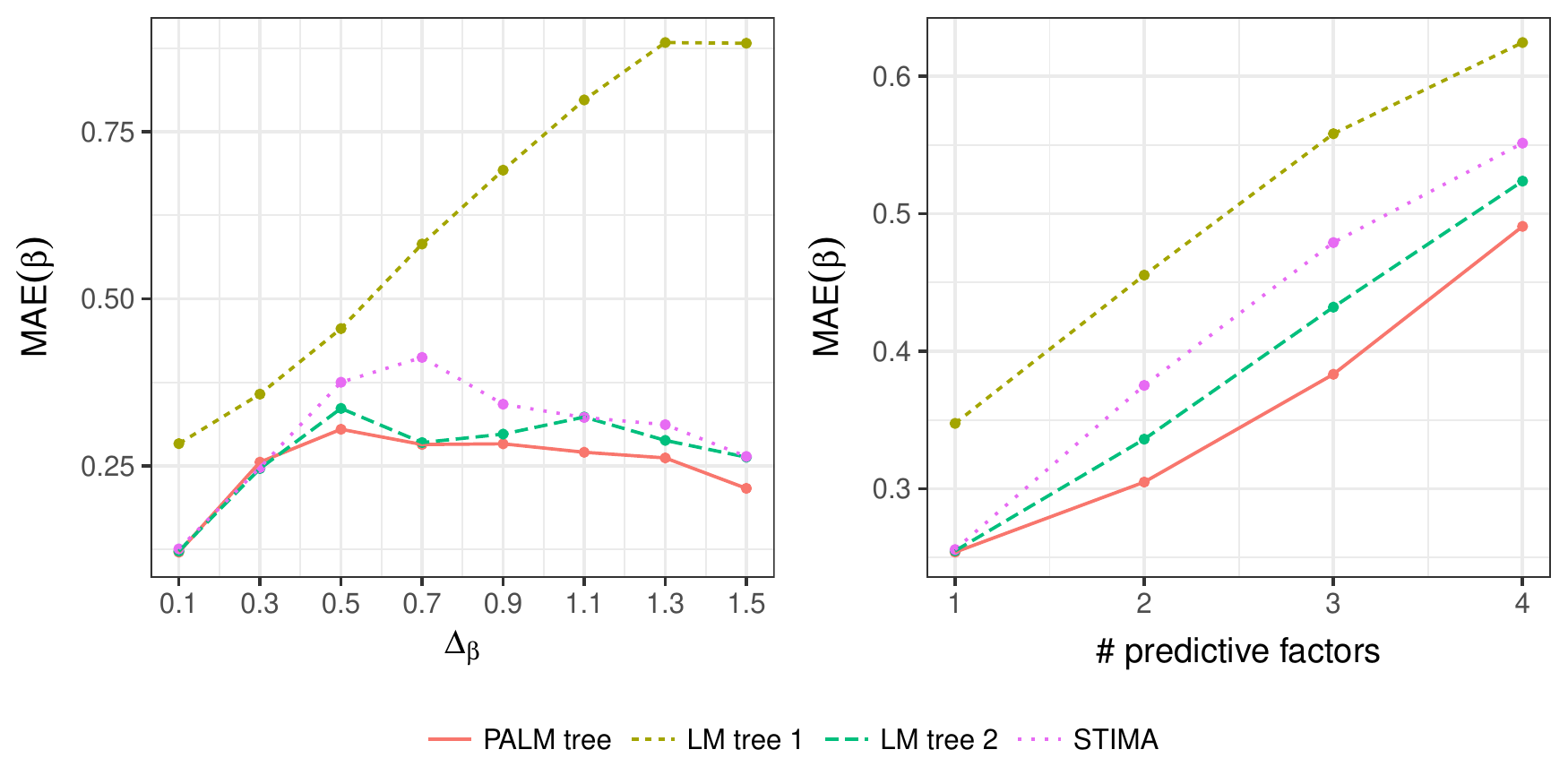}
\caption{Mean absolute difference between true and estimated treatment effect (mean absolute error, MAE; Question~\ref{sec:simest}).}\label{fig:fig_trteff}
\end{figure}

\end{knitrout}

Estimating or even predicting the correct treatment effect is the most
essential part of subgroup analysis. Even if one treatment is better
than the other, clinicians need to know if the difference is relevant.
\new{
 The evaluation of the treatment effect estimate can only be done for the
model-based recursive partitioning methods and STIMA since OTR is only designed
to produce binary decision rules.
}
The measure used to evaluate the treatment effect estimate is the mean absolute
difference between true and estimated treatment effect (mean absolute error,
MAE).  Figure~\ref{fig:fig_trteff} shows the MAE for the scenarios of varying
$\Delta_\beta$ and varying number of predictive factors. The error is smallest
for all three methods when the \emph{difference in treatment effect} is lowest
($\Delta_\beta = 0.1$), because even if the chosen subgroups are wrong, the
estimated treatment effect will likely be close to the true and very similar
treatment effects.
\new{
In this sense it is not a disadvantage that PALM tree, LM tree 2 and STIMA
often do not split into subgroups at all.  In fact, it may even be an
advantage, as the treatment effect estimate is then calculated based on a
larger data set and is less affected by random variability. The effect of the
small treatment difference gets less as the difference increases. However, as
the it increases, finding the correct subgroups becomes easier and the
error decreases.  At the same time finding the correct subgroups becomes easier
and slowly the error decreases again for PALM tree, LM tree 2 and STIMA.  For
this effect to be visible for LM tree 1, one would have to have larger
treatment effects, fewer prognostic factors and/or more observations, given the
large effect of the prognostic factor (see Figure~\ref{fig:fullf4} in the
Appedix). With an increasing \emph{number of predictive factors} the mean
absolute error in treatment effect increases.  The shape of the curve in
Figure~\ref{fig:fig_trteff} looks very different to the one in
Figure~\ref{fig:fig_cortrt}, even though they address similar questions, but
the more true predictive factors exist in the given simulation scenario the
harder it is for the methods to predict the treatment effect. This suggests
that simply knowing the more effective treatment does not tell the whole story.
This is supported across simulation scenarios (compare
Figures~\ref{fig:fullf2} and \ref{fig:fullf3}).
}

\section{Illustration: Treatment differences in mathematics exam}\label{sec:aplic}

The Mathematics 101 course for first-year business and economics students at
Universit\"at Innsbruck gives an introduction to mathematical analysis,
linear algebra, financial mathematics, and probability calculus. Students are
assessed by biweekly online tests during the semester and a written exam at the
end. The exam consists of 13 single-choice questions with 5 answer alternatives,
one of which is correct. Students who answer more than 60 percent of the
questions correctly pass the course. The percentage of successful online tests
captures math ability of the students and is a known predictor for success in
the final exam.

The data contains the exam results of 729 students (out of 941 who originally
registered for the course) for the fall semester in 2014/15. Due to limited
availability of seats in the exam room, the students were asked to select a
group, where the first group wrote the exam in the morning and the second group
right after the first group finished. The two groups received slightly
different questions on the same topics covering the scope of the course.  We
are interested in whether the exam is fair in the sense that it is on average
equally hard or difficult for the two groups.  In other words we want to find
out whether there is a ``treatment effect'' with the different selection of
exam questions in the two groups corresponding to the ``treatments''.  As a
first rather naive check we consider a simple one-way regression model for the
percentage of correct answers by group, as reported in the first column of
Table~\ref{tab:math}.  This yields an expected percentage of
57.6 for a student in group~1 and a
difference of 2.33 percentage points
for students in group~2. Thus, the model finds only a small drop in the
percentage of correctly solved answers and the corresponding confidence
interval includes a zero change.

However, in this first model we have neglected the influence of the students'
ability which is particularly relevant here because the students could freely
choose their exam group. Therefore, there might have been self-selection of
more (or less) able students into the first (or second) group. To account for
such ability effects in the model we include the percentage of points from the
previous online tests that captures the students' ability and preparation.
As shown in the second column of Table~\ref{tab:math} this variable is indeed
strongly associated with the exam results, where one additional percentage point
in the online tests leads to additional 0.86
expected percentage points in the written exam. More importantly, the group effect
increases to 4.37 and the corresponding confidence
interval does not include zero anymore. Despite the increase in the group
effect, the absolute size of the group difference is still moderate corresponding
to about half an exercise out of 13.

\begin{figure}[t!]
\includegraphics[width=1\textwidth]{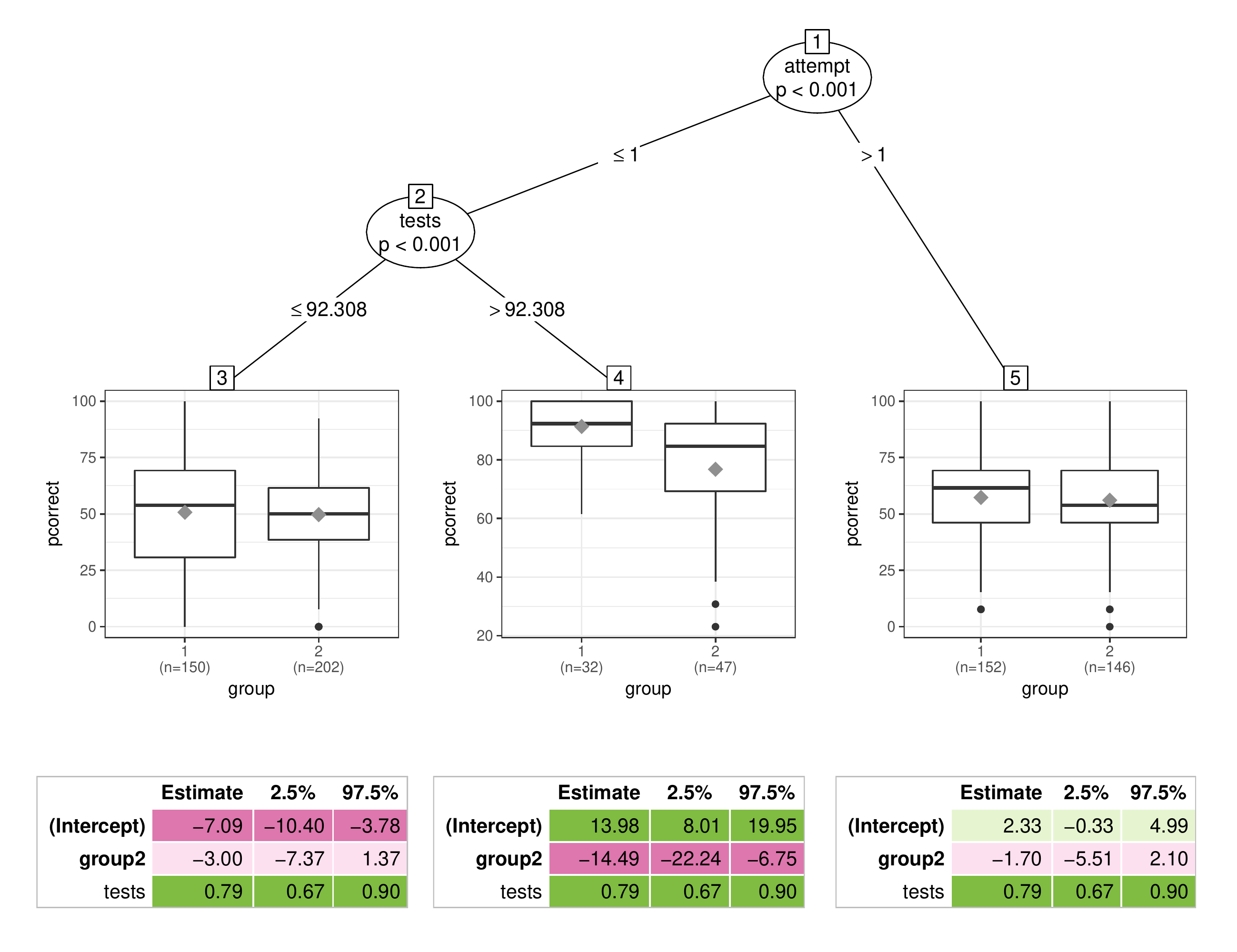}
\caption{PALM tree for the percentage of correct answers explained by
group differences while globally adjusting for ability (i.e., percentage of points
obtained in previous online tests).}\label{fig:palm}
\end{figure}

To explore the size of the treatment effect for the group differences further,
we consider the possibility that this may vary across subgroups of students.
Known student characteristics that may lead to such subgroups here are
gender, the number of semesters the student has already been studying, the
number of times the student has already attempted the exam, the type of study
(three year bachelor program vs.\ four year diploma program) and also the
ability/preparation as captured by percentage of successful exercises in the
online tests.
\nnew{
Since the test results in the online tests during the semester are known to
have an important direct effect on the performance in the exam, the test
parameter is included in the PALM tree.
}
Figure~\ref{fig:palm} shows the resulting PALM tree with the segmented local
group effect while adjusting for a global online tests effect.  The strongest
parameter instability is associated with the number of attempts and the group
of students in the first attempt are split a second time by the percentage from
the online tests. Two of the resulting subgroups (node~3 and~5) exhibit only
very small group differences but in node~4 the second group obtained clearly a
lower response percentage. This node is the smallest subgroup found and
encompasses the highly able students taking the course for the first time. For
this subsample the treatment effect is about 14 percentage points, which means
that the students in the second batch solved about two exercises less than
those in the first batch.

Overall this clearly conveys the strength of the PALM tree method: Especially
in situations where the coefficient of interest is modest in a main-effects
model and where further covariates are available whose influence on the main
model parameters is not obvious, the PALM tree is an attractive option to
globally control for certain variables while searching for local effects in others.
Note, however, that due to the forward selection of models/effects the
resulting confidence intervals in the terminal nodes (Table~\ref{tab:math}
and Figure~\ref{fig:palm}) should not be used for inference but
interpreted as a measure of variability.

\begin{table}
\centering
\begin{knitrout}
\definecolor{shadecolor}{rgb}{0.969, 0.969, 0.969}\color{fgcolor}
\includegraphics[width=1\textwidth]{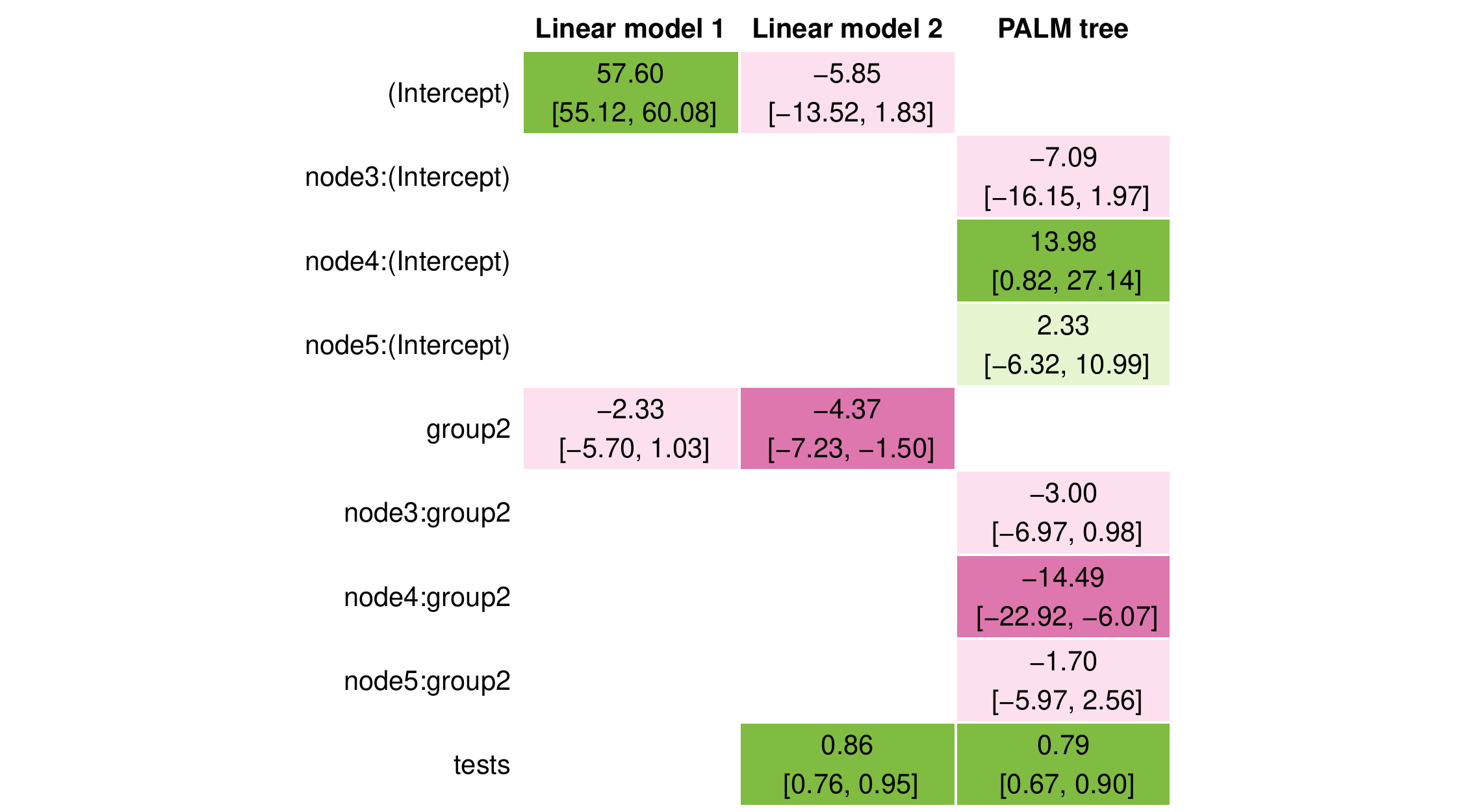}
\end{knitrout}
\caption{Three models for the mathematics exam data. The response variable is
the percentage of correctly solved exercises and the main covariat of interest
are the treatment differences between the first and second exam group.
Confidence intervals are given in brackets.}
\label{tab:math}
\end{table}

\section{Discussion}

Model-based trees are effective tools to identify subgroups in data which
differ in terms of model parameters. PALM trees are special model-based trees
where some parameters can be fixed globally for the entire sample and do not
depend on the subgroup structure. Our simulation study has shown that in cases
where there are such specified factors with a direct effect on the outcome,
PALM trees reliably detect the correct subgroups while at the same time having
a low probability of detecting subgroups when there are none.
\new{
STIMA is a flexible and well performing competitor of model-based trees.  The
most important downside of STIMA is that it is very slow with in some instances
single trees taking hours to compute (see Appendix~\ref{sec:ct}).  Moreover, it
has to be taken with a grain of salt that the R package ``stima'' is not
actively maintained on the Comprehensive R Archive Network.
}
Although optimal treatment regimes (OTR) perform comparably to PALM trees in
terms of detecting the best treatment option in the given simulation study,
PALM trees are typically better at recovering a parsimonious tree capturing the
underlying subgroup structure.  This makes PALM tree results easier to
interpret and to communicate to practitioners, which we believe is an important
advantage in many applications.  Moreover, the simulation study clearly showed
the effect of misspecifications in global vs.\ local effects in PALM trees.
While it is important to correctly identify the variables with \emph{additive}
effects (LM tree 1 vs.\ LM tree 2 or PALM tree), it is not so important to
correctly identify whether these additive effects are \emph{global or local}
(LM tree 2 vs.\ PALM tree).
\nnew{
However, by reducing the number of tests in the split procedure and focusing
only on certain relevant model parameters, some power and efficiency can be
gained from selecting a suitable PALM tree.
}

PALM trees allow exploring and questioning results of (generalised) linear
models.  The PALM tree analysis of the Mathematics 101 exam showed that a
linear model regressing the percentage points of correct anwers on the group
and earlier test results is too simple. Only for a relatively small subgroup of
students who attempted the exam for the first time and who showed good
performance during the semester it did make a difference whether they attempted
the exam in the first or second group.

Although large parts of this manuscript focus on subgroup analyses in clinical
trials, PALM trees can also be applied in a wide range of other applications as
well -- e.g., in the social sciences as shown in the mathematics exam
application case study.

\section*{Computational details}

Open-source implementations of the model-based tree algorithms LM tree and GLM
tree are available in the \textbf{partykit} package \citep[][functions
\code{lmtree()} and \code{glmtree()}]{hothorn_partykit_2015}. The PALM tree
algorithm is available in the \textbf{palmtree} package \citep[][function
\code{palmtree()}]{zeileis_palmtree_2017}. OTR is available in package
DynTxRegime \citep{holloway_r_2015}. The STIMA implementation has been archived
on CRAN but can still be downloaded from
\url{https://cran.r-project.org/src/contrib/Archive/stima/}. Simulations were
conducted using the batchtools package \citep{lang_batchtools_2017}.

The manuscript including simulation study and application can be reproduced
using the supplementary online material.

\section*{Acknowledgements}
We thank Andrea Farnham for improving the language. We are thankful to the
Swiss National Fund for funding this project with grants 205321\_163456 and
IZSEZ0\_177091 and mobility grant 205321\_163456/2.

\begin{appendix}

\section{Full factorial simulation}\label{sec:fullf}

\new{

The simulation study described in Section~\ref{sec:sim} takes a \emph{ceteris
paribus} approach and varies one simulation variable at a time while keeping
the others at a \emph{standard value}. We did an additional simulation study
where we vary all variables, which leads to $8 \cdot 5 \cdot 2 \cdot 4 \cdot 4
\cdot 4 = 5120$ (see Table~\ref{tab:sim}) different scenarios.  For each
scenario we simulated two data sets and ran all algorithms on each.
\nnew{
In the following we show a small selection of interesting graphics based on the
simulations. For the full results of the simulation studies we refer to the
online material.
}

Figure~\ref{fig:fullf1} shows the marginal results of the ARI for
$\Delta_\beta$, the number of predictive factors, the number of observations
and quantitative versus qualitative interactions. We average over the other
simulation variables and the two repetitions. For sake of easy visualisation,
we restrict the plotted variable to few levels. Similarly
Figures~\ref{fig:fullf2} and \ref{fig:fullf3} show the marginal results of the
proportion of correct treatment assignment and mean absolute error in estimated
treatment effect for the number of predictive factors, $\Delta_\beta$, the
number of observations and quantitative versus qualitative interactions.
Figure~\ref{fig:fullf4} shows the results for the MAE for $n = 900$ and one
prognostic factor to show when LM tree 1 starts to improve (see
Section~\ref{sec:simest}).

Figure~\ref{fig:fullf1} shows that PALM tree can handle simple subgroups with
one predictive factor even when the number of observations is low, but the
difference in treatment effects must be reasonably high.  All other algorithms
perform worse, with LM tree 2 and STIMA being the strongest competitors in the
low-n-scenarios. OTR performs reasonably well if qualitative subgroups are
present.  For $n = 500$ the performance of PALM tree rises already at lower
levels of $\Delta_\beta$. The performance of PALM tree and LM tree 2 is very
similar and STIMA also performs well. By design OTR ignores any non-qualitative
subgroups.

When quantitative treatment subgroups exist, all methods are good at deciding
the correct treatment regime (see Figure~\ref{fig:fullf2}), especially when the
number of observations is reasonably high (300). With $n = 100$ PALM tree, LM
tree 2, STIMA and even LM tree 1 still perform very well. OTR is the weakest
competitor here. With low numbers of observations ($n = 100$), low treatment
effect differences ($\Delta_\beta = 0.5$) and qualitative differences, the
performance of all algorithms is close to random guessing (0.5), irrespective
of the number of predictive factors. With higher $\Delta_\beta$ PALM tree
performs reasonably well, followed by LM tree 2, STIMA and OTR (order depending
on the number of predictive factors). For $n = 300$ and $\Delta_\beta = 0.5$
STIMA and LM tree 1 perform worst, but STIMA catches up with the other
algorithms when $\Delta_\beta = 1.5$, whereas LM tree 1 stays at the bottom.
Section~\ref{sec:simotr} discusses these results in the context of the results
in the star-like simulation study.

Section~\ref{sec:simest} already partly discussed Figures~\ref{fig:fullf3} and
\ref{fig:fullf4}. Figure~\ref{fig:fullf3} shows that across different
scenarios the MAE increases with increasing number of predictive factors.
PALM tree is among the best performers everywhere. In comparison to the
other algorithms it performs particularly well in low-n-qualitative scenarios
whith $\Delta_\beta = 1.5$.

\begin{knitrout}
\definecolor{shadecolor}{rgb}{0.969, 0.969, 0.969}\color{fgcolor}\begin{figure}
\includegraphics[width=\maxwidth]{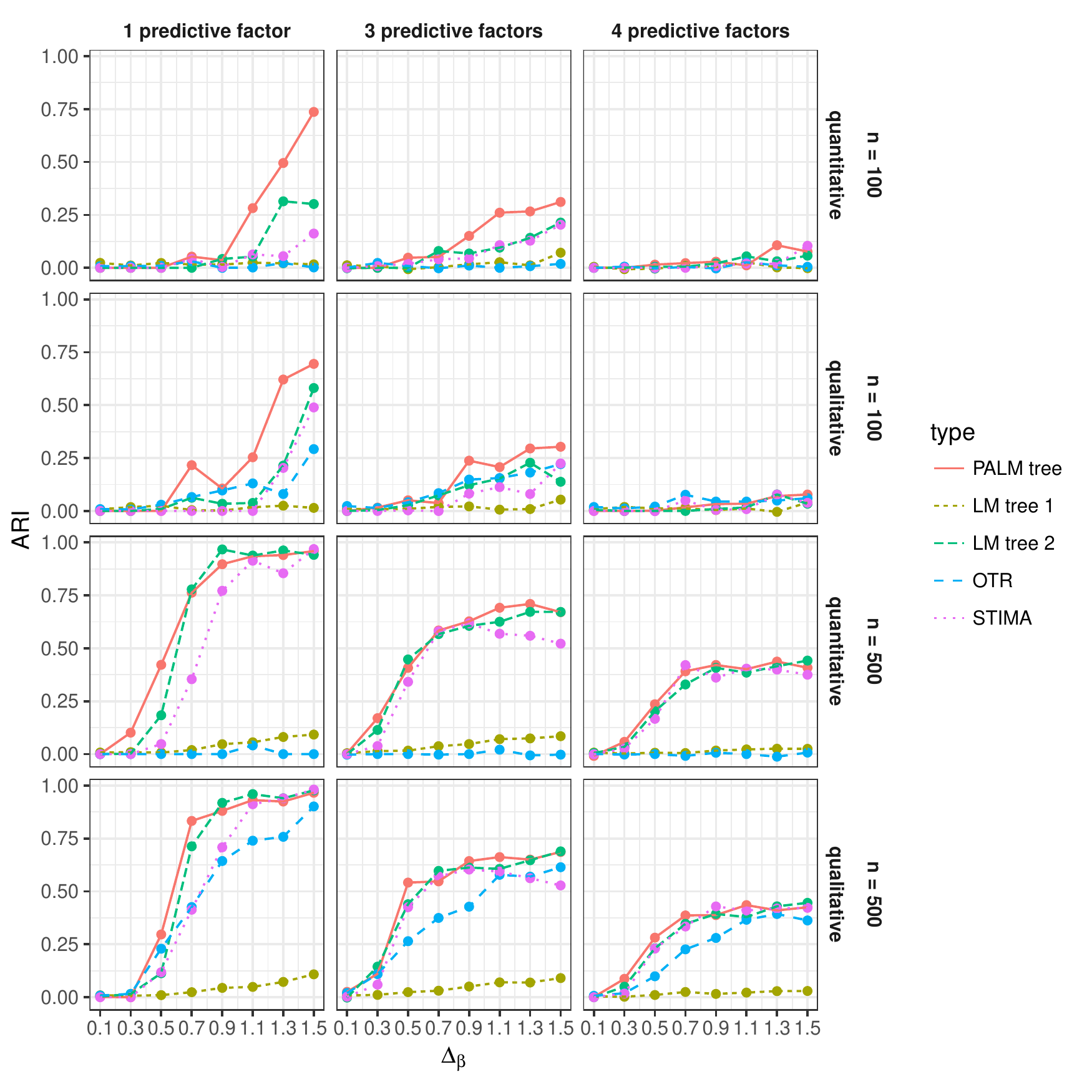}
\caption[Mean ARI in the full factorial design with two simulated data sets per design (Question 3.1)]{Mean ARI in the full factorial design with two simulated data sets per design (Question 3.1).}\label{fig:fullf1}
\end{figure}

\end{knitrout}

\begin{knitrout}
\definecolor{shadecolor}{rgb}{0.969, 0.969, 0.969}\color{fgcolor}\begin{figure}
\includegraphics[width=\maxwidth]{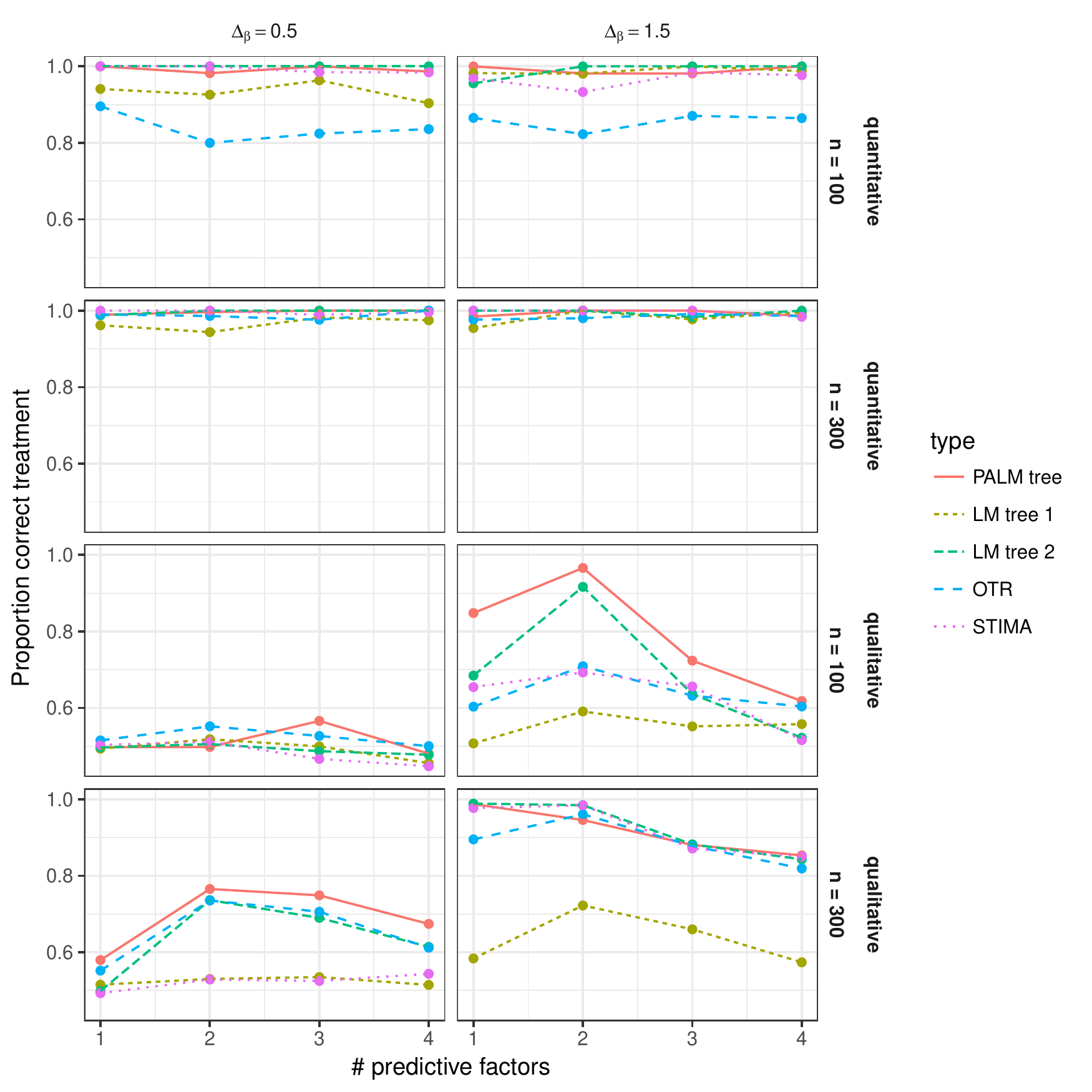}
\caption{Proportion of observations in all trees where better treatment is correctly identified in the full factorial design with two simulated data sets per design (Question~\ref{sec:simotr}).}\label{fig:fullf2}
\end{figure}

\end{knitrout}

\begin{knitrout}
\definecolor{shadecolor}{rgb}{0.969, 0.969, 0.969}\color{fgcolor}\begin{figure}
\includegraphics[width=\maxwidth]{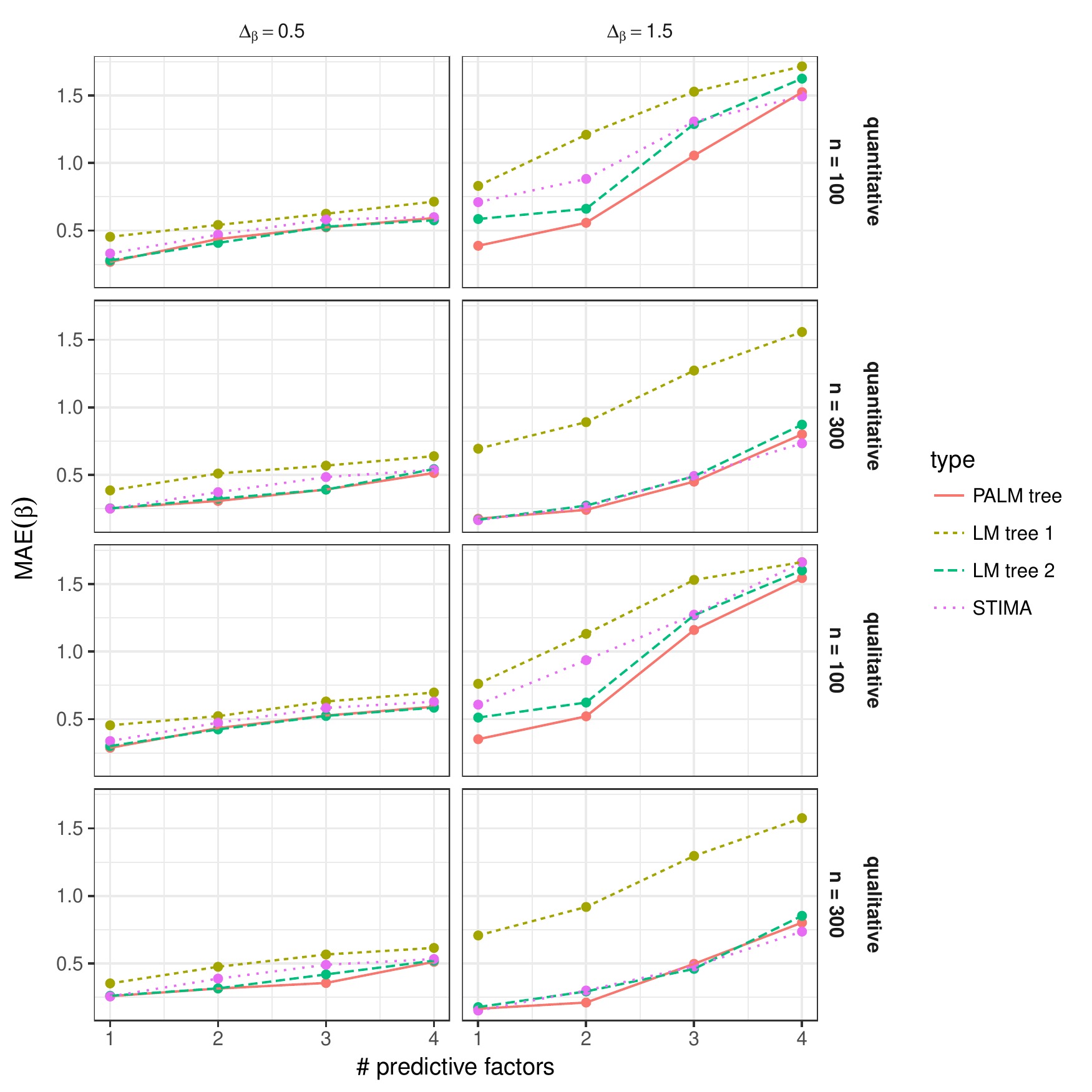}
\caption{Mean absolute difference between true and estimated treatment effect (mean absolute error, MAE) in the full factorial design with two simulated data sets per design (Question~\ref{sec:simest}).}\label{fig:fullf3}
\end{figure}

\end{knitrout}

\begin{knitrout}
\definecolor{shadecolor}{rgb}{0.969, 0.969, 0.969}\color{fgcolor}\begin{figure}

{\centering \includegraphics[width=0.65\textwidth]{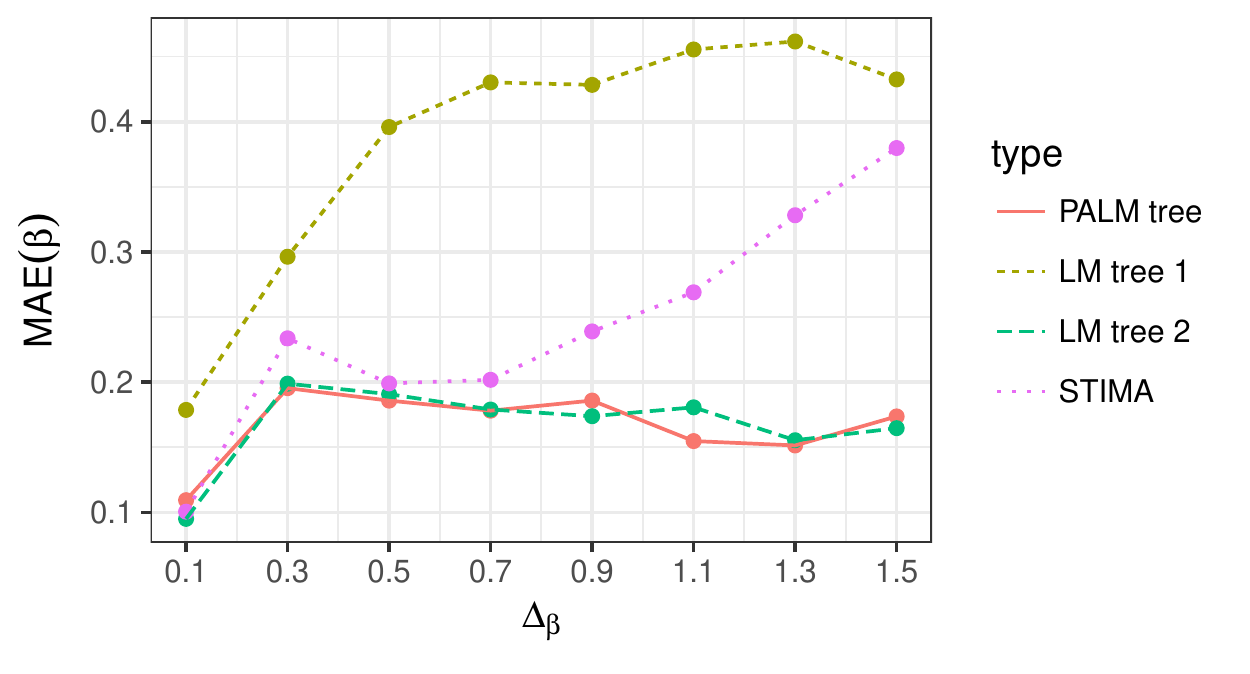}}
\caption{Mean absolute difference between true and estimated treatment effect (mean absolute error, MAE) in the full factorial design with two simulated data sets per design (Question~\ref{sec:simest}). Limited data to scenarios with 900 observations and one prognostic factor.}\label{fig:fullf4}
\end{figure}

\end{knitrout}

\section{Computation times}\label{sec:ct}

The computation times for all methods except STIMA are very reasonable in these
applications.  For a summary of computation times in the full factorial desing
see Table~\ref{ct}.  STIMA reached a maximum of 17.4 hours
and almost half the models took half an hour or longer.
\begin{table}[h]
\caption{Quantiles of computation times per algorithm in seconds.\label{ct}}
\begin{center}
\begin{tabular}{llllll}
\hline
\multicolumn{1}{c}{}&\multicolumn{1}{c}{0\%}&\multicolumn{1}{c}{25\%}&\multicolumn{1}{c}{50\%}&\multicolumn{1}{c}{75\%}&\multicolumn{1}{c}{100\%}\tabularnewline
\hline
PALM tree &0&0&1&2&7\tabularnewline
LM tree 1 &0&1&1&2&5\tabularnewline
LM tree 2 &0&0&1&1&4\tabularnewline
OTR &0&0&1&1&2\tabularnewline
STIMA&3&233.5&1941&8646.5&62512\tabularnewline
\hline
\end{tabular}\end{center}
\end{table}

}
\end{appendix}
\end{document}